\PassOptionsToPackage{unicode}{hyperref}
\PassOptionsToPackage{hyphens}{url}
\PassOptionsToPackage{dvipsnames,svgnames,x11names}{xcolor}
\documentclass[
]{article}
\usepackage{amsmath,amssymb}
\usepackage{lmodern}
\usepackage{iftex}
\ifPDFTeX
  \usepackage[T1]{fontenc}
  \usepackage[utf8]{inputenc}
  \usepackage{textcomp} % provide euro and other symbols
\else % if luatex or xetex
  \usepackage{unicode-math}
  \defaultfontfeatures{Scale=MatchLowercase}
  \defaultfontfeatures[\rmfamily]{Ligatures=TeX,Scale=1}
\fi
\IfFileExists{upquote.sty}{\usepackage{upquote}}{}
\IfFileExists{microtype.sty}{% use microtype if available
  \usepackage[]{microtype}
  \UseMicrotypeSet[protrusion]{basicmath} % disable protrusion for tt fonts
}{}
\makeatletter
\@ifundefined{KOMAClassName}{% if non-KOMA class
  \IfFileExists{parskip.sty}{%
    \usepackage{parskip}
  }{% else
    \setlength{\parindent}{0pt}
    \setlength{\parskip}{6pt plus 2pt minus 1pt}}
}{% if KOMA class
  \KOMAoptions{parskip=half}}
\makeatother
\usepackage{xcolor}
\usepackage{graphicx}
\makeatletter
\def\maxwidth{\ifdim\Gin@nat@width>\linewidth\linewidth\else\Gin@nat@width\fi}
\def\maxheight{\ifdim\Gin@nat@height>\textheight\textheight\else\Gin@nat@height\fi}
\makeatother
\setkeys{Gin}{width=\maxwidth,height=\maxheight,keepaspectratio}
\makeatletter
\def\fps@figure{htbp}
\makeatother
\NewDocumentCommand\citeproctext{}{}
\NewDocumentCommand\citeproc{mm}{%
  \begingroup\def\citeproctext{#2}\cite{#1}\endgroup}
\makeatletter
 \let\@cite@ofmt\@firstofone
 \def\@biblabel#1{}
 \def\@cite#1#2{{#1\if@tempswa , #2\fi}}
\makeatother
\newlength{\cslhangindent}
\newlength{\csllabelwidth}
\newenvironment{CSLReferences}[2] % #1 hanging-indent, #2 entry-spacing
 {\begin{list}{}{%
  \setlength{\itemindent}{0pt}
  \setlength{\leftmargin}{0pt}
  \setlength{\parsep}{0pt}
  \ifodd #1
   \setlength{\leftmargin}{\cslhangindent}
   \setlength{\itemindent}{-1\cslhangindent}
  \fi
  \setlength{\itemsep}{#2\baselineskip}}}
 {\end{list}}
\usepackage{calc}

\ifLuaTeX
\usepackage[bidi=basic]{babel}
\else
\usepackage[bidi=default]{babel}
\fi
\babelprovide[main,import]{american}
\def\languageshorthands#1{}
\ifLuaTeX
  \usepackage{selnolig}  % disable illegal ligatures
\fi
\IfFileExists{bookmark.sty}{\usepackage{bookmark}}{\usepackage{hyperref}}
\IfFileExists{xurl.sty}{\usepackage{xurl}}{} % add URL line breaks if available
\hypersetup{
  pdftitle={PPanGGOLiN V2: technical enhancement and extended
functionalities for prokaryotic pangenome analysis},
  pdfauthor={Jérôme Arnoux, Jean Mainguy, Adelme Bazin, Guillaume
Gautreau, Téo Lemane, David Vallenet, Alexandra Calteau},
  pdflang={en-US},
  colorlinks=true,
  linkcolor={Maroon},
  filecolor={Maroon},
  citecolor={Blue},
  urlcolor={Blue},
  pdfcreator={LaTeX via pandoc}}

\title{PPanGGOLiN V2: technical enhancement and extended functionalities
for prokaryotic pangenome analysis}

\definecolor{c53baa1}{RGB}{83,186,161}
\definecolor{c202826}{RGB}{32,40,38}
\def \rorglobalscale {0.1}
\newcommand{\rorlogo}{%
\begin{tikzpicture}[y=1cm, x=1cm, yscale=\rorglobalscale,xscale=\rorglobalscale, every node/.append style={scale=\rorglobalscale}, inner sep=0pt, outer sep=0pt]
  \begin{scope}[even odd rule,line join=round,miter limit=2.0,shift={(-0.025, 0.0216)}]
    \path[fill=c53baa1,nonzero rule,line join=round,miter limit=2.0] (1.8164, 3.012) -- (1.4954, 2.5204) -- (1.1742, 3.012) -- (1.8164, 3.012) -- cycle;
    \path[fill=c53baa1,nonzero rule,line join=round,miter limit=2.0] (3.1594, 3.012) -- (2.8385, 2.5204) -- (2.5172, 3.012) -- (3.1594, 3.012) -- cycle;
    \path[fill=c53baa1,nonzero rule,line join=round,miter limit=2.0] (1.1742, 0.0669) -- (1.4954, 0.5588) -- (1.8164, 0.0669) -- (1.1742, 0.0669) -- cycle;
    \path[fill=c53baa1,nonzero rule,line join=round,miter limit=2.0] (2.5172, 0.0669) -- (2.8385, 0.5588) -- (3.1594, 0.0669) -- (2.5172, 0.0669) -- cycle;
    \path[fill=c202826,nonzero rule,line join=round,miter limit=2.0] (3.8505, 1.4364).. controls (3.9643, 1.4576) and (4.0508, 1.5081) .. (4.1098, 1.5878).. controls (4.169, 1.6674) and (4.1984, 1.7642) .. (4.1984, 1.8777).. controls (4.1984, 1.9719) and (4.182, 2.0503) .. (4.1495, 2.1132).. controls (4.1169, 2.1762) and (4.0727, 2.2262) .. (4.0174, 2.2635).. controls (3.9621, 2.3006) and (3.8976, 2.3273) .. (3.824, 2.3432).. controls (3.7505, 2.359) and (3.6727, 2.367) .. (3.5909, 2.367) -- (2.9676, 2.367) -- (2.9676, 1.8688).. controls (2.9625, 1.8833) and (2.9572, 1.8976) .. (2.9514, 1.9119).. controls (2.9083, 2.0164) and (2.848, 2.1056) .. (2.7705, 2.1791).. controls (2.6929, 2.2527) and (2.6014, 2.3093) .. (2.495, 2.3487).. controls (2.3889, 2.3881) and (2.2728, 2.408) .. (2.1468, 2.408).. controls (2.0209, 2.408) and (1.905, 2.3881) .. (1.7986, 2.3487).. controls (1.6925, 2.3093) and (1.6007, 2.2527) .. (1.5232, 2.1791).. controls (1.4539, 2.1132) and (1.3983, 2.0346) .. (1.3565, 1.9436).. controls (1.3504, 2.009) and (1.3351, 2.0656) .. (1.3105, 2.1132).. controls (1.2779, 2.1762) and (1.2338, 2.2262) .. (1.1785, 2.2635).. controls (1.1232, 2.3006) and (1.0586, 2.3273) .. (0.985, 2.3432).. controls (0.9115, 2.359) and (0.8337, 2.367) .. (0.7519, 2.367) -- (0.1289, 2.367) -- (0.1289, 0.7562) -- (0.4837, 0.7562) -- (0.4837, 1.4002) -- (0.6588, 1.4002) -- (0.9956, 0.7562) -- (1.4211, 0.7562) -- (1.0118, 1.4364).. controls (1.1255, 1.4576) and (1.2121, 1.5081) .. (1.2711, 1.5878).. controls (1.2737, 1.5915) and (1.2761, 1.5954) .. (1.2787, 1.5991).. controls (1.2782, 1.5867) and (1.2779, 1.5743) .. (1.2779, 1.5616).. controls (1.2779, 1.4327) and (1.2996, 1.3158) .. (1.3428, 1.2113).. controls (1.3859, 1.1068) and (1.4462, 1.0176) .. (1.5237, 0.944).. controls (1.601, 0.8705) and (1.6928, 0.8139) .. (1.7992, 0.7744).. controls (1.9053, 0.735) and (2.0214, 0.7152) .. (2.1474, 0.7152).. controls (2.2733, 0.7152) and (2.3892, 0.735) .. (2.4956, 0.7744).. controls (2.6016, 0.8139) and (2.6935, 0.8705) .. (2.771, 0.944).. controls (2.8482, 1.0176) and (2.9086, 1.1068) .. (2.952, 1.2113).. controls (2.9578, 1.2253) and (2.9631, 1.2398) .. (2.9681, 1.2544) -- (2.9681, 0.7562) -- (3.3229, 0.7562) -- (3.3229, 1.4002) -- (3.4981, 1.4002) -- (3.8349, 0.7562) -- (4.2603, 0.7562) -- (3.8505, 1.4364) -- cycle(0.9628, 1.7777).. controls (0.9438, 1.7534) and (0.92, 1.7357) .. (0.8911, 1.7243).. controls (0.8623, 1.7129) and (0.83, 1.706) .. (0.7945, 1.7039).. controls (0.7588, 1.7015) and (0.7252, 1.7005) .. (0.6932, 1.7005) -- (0.4839, 1.7005) -- (0.4839, 2.0667) -- (0.716, 2.0667).. controls (0.7477, 2.0667) and (0.7805, 2.0643) .. (0.8139, 2.0598).. controls (0.8472, 2.0553) and (0.8768, 2.0466) .. (0.9025, 2.0336).. controls (0.9282, 2.0206) and (0.9496, 2.0021) .. (0.9663, 1.9778).. controls (0.9829, 1.9534) and (0.9914, 1.9209) .. (0.9914, 1.8799).. controls (0.9914, 1.8362) and (0.9819, 1.8021) .. (0.9628, 1.7777) -- cycle(2.6125, 1.3533).. controls (2.5889, 1.2904) and (2.5553, 1.2359) .. (2.5112, 1.1896).. controls (2.4672, 1.1433) and (2.4146, 1.1073) .. (2.3529, 1.0814).. controls (2.2916, 1.0554) and (2.2228, 1.0427) .. (2.1471, 1.0427).. controls (2.0712, 1.0427) and (2.0026, 1.0557) .. (1.9412, 1.0814).. controls (1.8799, 1.107) and (1.8272, 1.1433) .. (1.783, 1.1896).. controls (1.7391, 1.2359) and (1.7052, 1.2904) .. (1.6817, 1.3533).. controls (1.6581, 1.4163) and (1.6465, 1.4856) .. (1.6465, 1.5616).. controls (1.6465, 1.6359) and (1.6581, 1.705) .. (1.6817, 1.7687).. controls (1.7052, 1.8325) and (1.7388, 1.8873) .. (1.783, 1.9336).. controls (1.8269, 1.9799) and (1.8796, 2.0159) .. (1.9412, 2.0418).. controls (2.0026, 2.0675) and (2.0712, 2.0804) .. (2.1471, 2.0804).. controls (2.223, 2.0804) and (2.2916, 2.0675) .. (2.3529, 2.0418).. controls (2.4143, 2.0161) and (2.467, 1.9799) .. (2.5112, 1.9336).. controls (2.5551, 1.8873) and (2.5889, 1.8322) .. (2.6125, 1.7687).. controls (2.636, 1.705) and (2.6477, 1.6359) .. (2.6477, 1.5616).. controls (2.6477, 1.4856) and (2.636, 1.4163) .. (2.6125, 1.3533) -- cycle(3.8015, 1.7777).. controls (3.7825, 1.7534) and (3.7587, 1.7357) .. (3.7298, 1.7243).. controls (3.701, 1.7129) and (3.6687, 1.706) .. (3.6333, 1.7039).. controls (3.5975, 1.7015) and (3.5639, 1.7005) .. (3.5319, 1.7005) -- (3.3226, 1.7005) -- (3.3226, 2.0667) -- (3.5547, 2.0667).. controls (3.5864, 2.0667) and (3.6192, 2.0643) .. (3.6526, 2.0598).. controls (3.6859, 2.0553) and (3.7155, 2.0466) .. (3.7412, 2.0336).. controls (3.7669, 2.0206) and (3.7883, 2.0021) .. (3.805, 1.9778).. controls (3.8216, 1.9534) and (3.8301, 1.9209) .. (3.8301, 1.8799).. controls (3.8301, 1.8362) and (3.8206, 1.8021) .. (3.8015, 1.7777) -- cycle;
  \end{scope}
\end{tikzpicture}
}

\usepackage[affil-it]{authblk}
\usepackage{orcidlink}
\author[1%
  *%
  ]{Jérôme Arnoux%
    \,\orcidlink{0000-0003-2769-3006}\,%
    }
\author[1%
  *%
  ]{Jean Mainguy%
    \,\orcidlink{0009-0006-9160-9744}\,%
    }
\author[2%
  ]{Adelme Bazin%
    \,\orcidlink{0000-0002-5656-4708}\,%
    }
\author[3%
  ]{Guillaume Gautreau%
    \,\orcidlink{0000-0002-0970-9361}\,%
    }
\author[1%
  ]{Téo Lemane%
    \,\orcidlink{0000-0002-7210-3178}\,%
    }
\author[1%
  ]{David Vallenet%
    \,\orcidlink{0000-0001-6648-0332}\,%
    }
\author[1%
  \ensuremath\mathparagraph]{Alexandra Calteau%
    \,\orcidlink{0000-0002-5871-9347}\,%
    }

\affil[1]{LABGeM, Genomique Métabolique, CEA, Genoscope, Institut
François Jacob, Université d'Évry, Université Paris-Saclay, CNRS,
France.%
    \,\protect\href{https://ror.org/00xc55v17}{\protect\rorlogo}\,%
  }
\affil[2]{Laboratory of Biology and Modeling of the Cell, Ecole Normale
Supérieure de Lyon, CNRS, Université Claude Bernard Lyon, Université de
Lyon, France%
    \,\protect\href{https://ror.org/01bj4fd12}{\protect\rorlogo}\,%
  }
\affil[3]{Université Paris-Saclay, INRAE, MaIAGE, Jouy-en-Josas, France%
    \,\protect\href{https://ror.org/05qdnns64}{\protect\rorlogo}\,%
  }
\affil[$\mathparagraph$]{Corresponding author: %
}
\affil[*]{These authors contributed equally.}
\date{23 June 2026}

\begin{document}
\maketitle

\section{Summary}\label{summary}

The exponential growth of genomic data, particularly for microbes, has
made pangenomic approaches a gold standard for large-scale comparative
genomics. By capturing the full genomic diversity of a species rather
than relying on a single reference, pangenomics has transformed
microbial genomics, revealing the adaptive potential of bacteria and the
evolutionary dynamics underlying functional diversity. Among available
tools, PPanGGOLiN distinguishes itself through its graph-based model
coupled with statistical gene partitioning.

Here we present PPanGGOLiN v2, which introduces substantial improvements
across three dimensions: new analytical features that expand what users
can investigate, a comprehensive software architecture redesign that
improves maintainability and extensibility, and performance improvements
that address the computational demands of ever-growing genomic datasets.

\section{Statement of need}\label{statement-of-need}

In the last decade, the number of sequenced microbial genomes has
exploded from a few thousand to several millions. While this wealth of
data offers immense potential, traditional genome-centric approaches
reach their limits when exploring and interpreting variation at this
scale (\citeproc{ref-land_insights_2015}{Land et al., 2015};
\citeproc{ref-parks_standardized_2018}{Parks et al., 2018}). The
pangenome addresses this by condensing all genomic information from a
set of genomes into a single structure, capturing not only gene
presence/absence but also genetic context, e.g.~mobile genetic elements
or functional annotations, at the species or clade level
(\citeproc{ref-tettelin_genome_2005}{Tettelin et al., 2005}).

PPanGGOLiN (\citeproc{ref-gautreau_ppanggolin_2020}{Gautreau et al.,
2020}) addresses these needs through a graph-based model that clusters
genes into families based on sequence similarity and encodes
neighborhood information, enabling high compression of genomic diversity
into a single data structure. Gene families are then partitioned using a
Bernoulli Mixture Model coupled with a Markov Random Field that takes
into account both family occurrence and genomic neighborhood. The
software suite further integrates PanRGP
(\citeproc{ref-bazin_panrgp_2020}{Bazin et al., 2020}), for
identification of Regions of Genomic Plasticity (RGPs) as well as their
insertion spots, and PanModule
(\citeproc{ref-bazin_panmodule_2021}{Bazin et al., 2021}), for their
description as conserved modules.

PPanGGOLiN v2 builds on this foundation by adding new analytical
features: genomic context search, RGP clustering, pangenome projection
and metadata association (\autoref{fig:overview}) together with a full
architectural redesign and a modernized development workflow, making the
tool more capable, maintainable, and open to community contributions.

\section{State of the field}\label{state-of-the-field}

Pangenomics has transformed prokaryotic genome analyses by moving beyond
a mainly reference-centric approach toward new representations that
capture the entire genomic diversity of microbial populations. Two
representational models for pangenome graphs are widely used.
Sequence-level graphs, with tools such as Bifrost
(\citeproc{ref-holley_bifrost_2020}{Holley \& Melsted, 2020}) or the vg
toolkit (\citeproc{ref-garrison_variation_2018}{Garrison et al., 2018}),
encode variation at the nucleotide level, providing resolution of any
small variants, but at a high computational cost and interpretive
complexity. Gene-level graphs with tools such as Panaroo
(\citeproc{ref-tonkin-hill_producing_2020}{Tonkin-Hill et al., 2020}) or
pangene (\citeproc{ref-li_exploring_2024}{Li et al., 2024}) instead
cluster genes into families to form graph nodes, with edges encoding
genomic contiguity. This representation neglects fine-scale sequence
variation but remains highly informative about gene presence/absence and
mobile genetic elements at a fraction of the resource cost.

Most gene-level tools rely on a predefined occurrence threshold to split
gene families into different partitions, which leads to inconsistent
predictions depending on the quality and diversity of the available
genomes. PPanGGOLiN addresses this using a statistical method relying on
both the patterns of occurrence of gene families, and the pangenome
graph topology to partition gene families into: \emph{persistent},
\emph{shell}, and \emph{cloud} genomes. Among dozens of alternative
gene-level pangenomic tools, only mOTUpan
(\citeproc{ref-buck_motupan_2022}{Buck et al., 2022}) and micropan
(\citeproc{ref-snipen_micropan_2015}{Snipen \& Liland, 2015}) use
statistical methods to compute pangenome partitions.

Additionally, PPanGGOLiN builds on its graph-based representation and
fine-grained partitioning to enable the identification of regions of
genomic plasticity and of conserved modules --- analyses that remain
unique to PPanGGOLiN.

\section{New features}\label{new-features}

\subsection{Projection: pangenome annotation of external
genomes}\label{projection-pangenome-annotation-of-external-genomes}

The projection feature enables annotation of external genomes using a
previously computed pangenome, without recomputing the full pangenome
structure. Genes from the input genome are assigned to existing gene
families and partitions based on sequence similarity; genes without
matches are considered genome-specific and assigned to the cloud
partition. From these assignments, PPanGGOLiN further predicts Regions
of Genomic Plasticity (RGPs), insertion spots, and conserved modules
present in the projected genome. This is particularly useful for
comparing newly sequenced genomes against an established reference
pangenome.

In Version 2, PPanGGOLiN introduces a new output format compatible with
Proksee (\citeproc{ref-grant_proksee_2023}{Grant et al., 2023}),
enabling visualization of the circular genomes annotated together with
pangenome information either from the pangenome itself or via projection
allowing further interactive analyses directly within the Proksee
platform.

\subsection{Genomic context
extraction}\label{genomic-context-extraction}

A genomic context refers here to a set of genes conserved within similar
genomic regions across multiple genomes. The pangenome graph already
encodes neighborhood relationships and is therefore well suited for
extracting such context. As input, PPanGGOLiN takes sequences
corresponding to genes of interest, which are aligned to the reference
sequence of each gene family to identify the target families in the
pangenome. Once identified, gene families present within a defined
genomic window around the target are extracted to construct a modified
pangenome graph using a transitive closure taking into account a larger
neighborhood.

The complete method is described in detail by Arnoux \emph{
et al.}
in the PANORAMA paper (\citeproc{ref-arnoux_panorama_2026}{Arnoux et al., 2026}). This feature provides a species-level view of
conserved genomic neighborhoods, enabling users to identify functionally
related genes and, for example, accurately reconstruct metabolic
pathways.

\subsection{RGP clustering}\label{rgp-clustering}

A new feature allows the comparative analysis of RGPs across genomes
within the pangenome. Clustering is based on the similarity of gene
content, where two RGPs are considered related if their genes belong to
the same gene families. Similarity is quantified using a Gene Repertoire
Relatedness (GRR) score.

Clustering is performed through graph modeling: each RGP is represented
as a node, and an edge is added between two nodes if their GRR (min or
max) exceeds a defined threshold (0.8 by default). Each connected
component of the resulting graph corresponds to a cluster of RGPs
sharing a common gene repertoire. Results are exportable as a tabulated
file or in formats compatible with graph visualization tools such as
Gephi(\citeproc{ref-bastian_gephi_2009}{Bastian et al., 2009}) and
Cytoscape (\citeproc{ref-shannon_cytoscape_2003}{Shannon et al., 2003}).
This analysis facilitates the exploration of the diversity and
dissemination of mobile genetic elements across genomes.

\subsection{Metadata}\label{metadata}

Another key addition is the ability to associate metadata with any
pangenome element, including genes, contigs, genomes, gene families,
edges, RGPs, spots, and modules. Metadata is provided as a tabulated
file requiring at minimum the identifier of the target element, with no
restriction on the type or number of additional fields, making the
format highly flexible. Each metadata entry is linked to a named source,
allowing the same element to carry multiple annotations from different
sources simultaneously.

For performance and portability, metadata is stored directly in the
pangenome file. While not used directly in pangenome computation, they
are propagated to all PPanGGOLiN outputs, facilitating downstream
interpretation and exploration of results.

Combined with RGP clustering, and pangenome projection, metadata
association bridges the gap between pangenome structure and biological
interpretation, allowing users to directly connect genomic variation to
functional, ecological, or experimental data.

\section{Software design}\label{software-design}

\begin{figure}
\centering
\includegraphics[keepaspectratio]{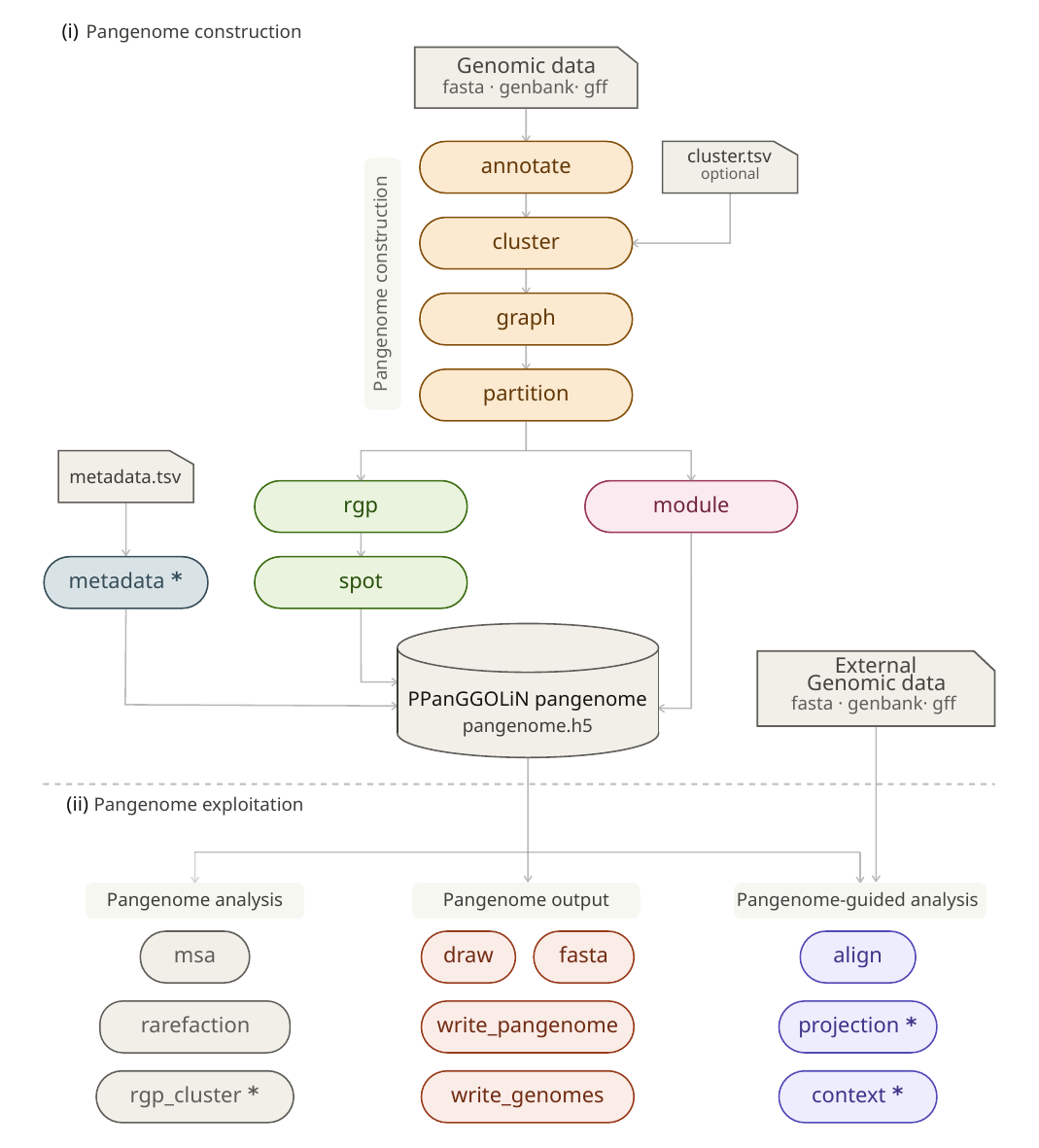}
\caption{\textbf{Overview of the PPanGGOLiN v2 workflow.} Each rounded
box represents a command of the software. Commands marked with * are new
in PPanGGOLiN v2. Commands are grouped into two sections. (\emph{i})
\emph{Pangenome construction}: starting from genomic data (with
annotations or not), the \texttt{annotate}, \texttt{cluster},
\texttt{graph}, and \texttt{partition} commands progressively build the
pangenome graph and partition gene families into \emph{persistent},
\emph{shell}, and \emph{cloud} components. Then, two independent
analyses can be performed: \texttt{rgp} followed by \texttt{spot}
identify Regions of Genomic Plasticity (RGPs) and their insertion spots,
while \texttt{module} identifies conserved genomic modules. All results
are stored in a single HDF5 file (\texttt{pangenome.h5}), which serves
as the central data object. Metadata from external sources can be
associated with any pangenome element via the \texttt{metadata} command
(*), using a tabulated input file. (\emph{ii}) \emph{Pangenome
exploitation}: the HDF5 file is used as input for four categories of
downstream commands. \emph{Pangenome analysis}: \texttt{msa} computes
multiple sequence alignments for gene families; \texttt{rarefaction}
computes the rarefaction curve of the pangenome; \texttt{rgp\_cluster}
(*) clusters RGPs based on shared gene content. \emph{Pangenome-guided
analysis}: \texttt{align} maps an external genome or protein set to
pangenome families; \texttt{context} (*) extracts conserved genomic
neighborhoods around genes of interest; \texttt{projection} (*)
annotates a new genome using an existing pangenome. \emph{Pangenome
output}: \texttt{draw}, \texttt{fasta}, \texttt{write\_pangenome}, and
\texttt{write\_genomes} export results in various formats for downstream
analyses and visualization.\label{fig:overview}}
\end{figure}

PPanGGOLiN is built around a central data model in which the HDF5
pangenome file serves as the reference data object for the entire
system. Analyses can be performed either step by step or through
workflow commands that execute the pipeline in a single run, providing
both flexibility and convenience (\autoref{fig:overview}). Once
executed, the HDF5 pangenome file supports multiple uses: exporting to
other file formats for downstream analyses, serving as input for
additional PPanGGOLiN commands, and being accessed programmatically
through the Python API in custom scripts. This central-object design
also enhances reproducibility: parameters are resolved consistently
(command line, then configuration file, then defaults), validated across
analysis steps, and stored within the pangenome file, enabling analyses
to be reliably reproduced and compared.

\section{Technical enhancements}\label{technical-enhancements}

PPanGGOLiN was originally developed as a research project. Despite
having robust and user-friendly workflows, its underlying codebase
required a complete redesign and refactoring to meet long-term software
engineering standards.

\begin{figure}
\centering
\includegraphics[keepaspectratio]{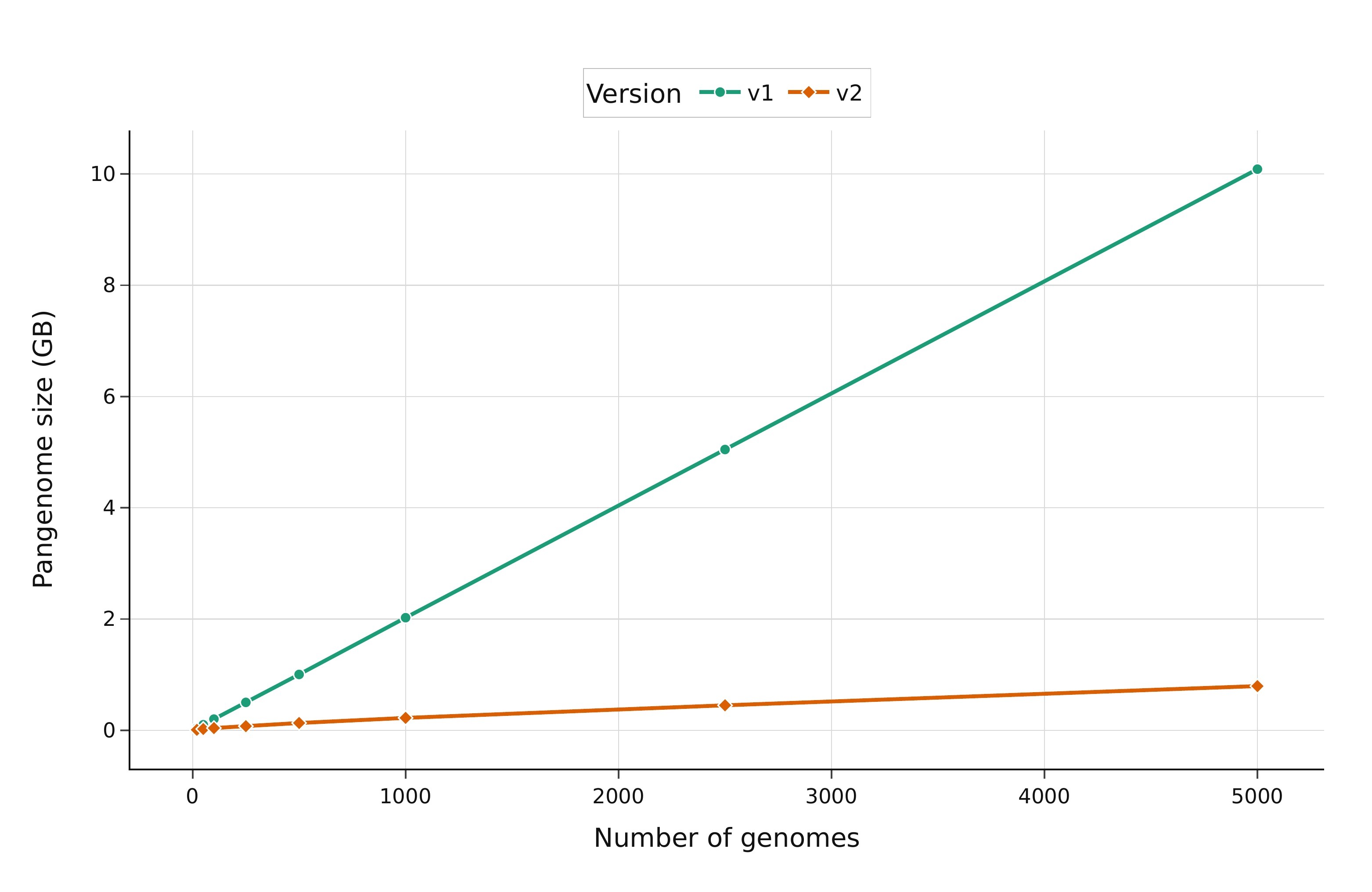}
\caption{\textbf{Pangenome HDF5 file size (GB) generated by PPanGGOLiN
v1.2.74 vs v2.3.0} \textbf{across increasing genome counts.} PPanGGOLiN
v2 consistently produces smaller files, supporting the impact of the v2
file structure redesign.\label{fig:hdf5_size}}
\end{figure}

Several targeted improvements address performance and reproducibility.
Prodigal (\citeproc{ref-hyatt_prodigal_2010}{Hyatt et al., 2010}) has
been replaced by Pyrodigal
(\citeproc{ref-larralde_pyrodigal_2022}{Larralde, 2022}), reducing I/O
overhead during genome annotation. The HDF5 pangenome file structure has
been redesigned to drastically reduce file size
(\autoref{fig:hdf5_size}) and improve programmatic access via API.
Pangenome file reading was optimized, reducing loading time, especially
for spots from over several minutes to few seconds for large datasets.
Memory consumption during sequence writing was reduced by reading
sequences directly from HDF5 files rather than loading them into memory.
A configuration file system has been introduced to facilitate the
integration of PPanGGOLiN into computational platforms (see Research
impact), and error management has been overhauled to provide clearer and
more informative user feedback.

Finally, documentation has been fully revised and is now hosted on
ReadTheDocs, lowering the barrier for new users. PPanGGOLiN v2 is
distributed via Bioconda(\citeproc{ref-gruning_bioconda_2018}{Grüning et
al., 2018}) and PyPI, ensuring broad accessibility across computing
environments.

\section{Research impact statement}\label{research-impact-statement}

PPanGGOLiN is widely adopted in microbial genomics, with
\textasciitilde300 citations on Google Scholar. Its results are
integrated into MicroScope
(\citeproc{ref-vallenet_microscope_2020}{Vallenet et al., 2020}), a
microbial genome annotation and analysis platform, and a dedicated
Galaxy (\citeproc{ref-the_galaxy_community_galaxy_2024}{The Galaxy
Community, 2024}) module has been developed to broaden accessibility.
The tool serves as a core dependency for PANORAMA
(\citeproc{ref-arnoux_panorama_2026}{Arnoux et al., 2026}),
enabling cross-species biological system predictions and pangenome
comparisons, and supports
\href{https://pangbank.genoscope.cns.fr/}{PanGBank}, a pangenome
database. A companion project further extends this ecosystem by
integrating PPanGGOLiN pangenomes --- including RGPs, spots, and modules
--- alongside functional metadata annotations into a Neo4j graph
database, enabling advanced graph-based queries across pangenome
elements (\citeproc{ref-arnoux_integrating_2024}{Arnoux et al., 2024}).
PPanGGOLiN is actively maintained and evolves through community
contributions, bug reports, and feature suggestions. In 2023, it
received the
\href{https://www.ouvrirlascience.fr/the-2023-open-science-free-research-software-awards/}{Open
Science Award for Open Source Research Software} from the French
Ministry of Higher Education and Research.

\section{AI usage disclosure}\label{ai-usage-disclosure}

AI tools were used in a limited capacity during the later stages of
development. In particular, GitHub Copilot (mainly using Claude Sonnet
4.6) was used for code review suggestions, assistance with writing some
tests, and documentation. Claude Sonnet 4.6 (Anthropic) was used to
assist in the design and iterative refinement of the workflow overview
figure (\autoref{fig:overview}) and manuscript proofreading. The core
software design, implementation, and architectural decisions were
developed independently without AI assistance. All AI-generated outputs
were systematically reviewed and validated by the authors.

\section{Acknowledgments}\label{acknowledgments}

This research was supported in part by the CFR PhD program of the French
Alternative Energies and Atomic Energy Commission (CEA) for JA and the
BlueRemediomics project for JM, which is funded by the European Union
under the Horizon Europe Program, Grant Agreement No.~101082304.

We are grateful to the PPanGGOLiN user community for their
contributions, feedback, and bug reports, which have continuously
improved the tool and shaped the development of this new version.

\section*{References}\label{references}
\addcontentsline{toc}{section}{References}

\protect\phantomsection\label{refs}

\end{document}